# Quantification of Tides in Giant Planets from Observations

Valéry Lainey[1] · Marco Zannoni[2] · Vincent Robert[1,3] · Tristan Guillot[4]

**Abstract**
Quantifying tidal effects on giant planets has recently made significant advances, thanks in particular to the Cassini space probe. During its thirteen-year orbit around Saturn, numerous measurements from different instruments made it possible to characterize fundamental parameters such as Saturn's Love number $k_2$ and quality factor Q at different frequencies. In this article, we summarize the various measurements and methods that have allowed to arrive at such a result, as well as the extrapolations that can be deduced for other systems. More generally, the state of the art concerning the four giant planets of the Solar System is presented, as well as the case of exoplanets.

## 1 Introduction

Knowledge of moon dynamics has significantly improve over the last 150 years. The advent of new, ever more precise instruments, including heliometers and the use of photographic plates, meant that by the end of the 19th century, precision had rapidly increased from 1 arcsecond to 200 milliarcsecond (mas). These observations were then used by analytical theories to produce precise ephemerides for these moons. In the case of the Jovian system, Sampson's analytical theory (Sampson 1921) was the reference throughout the first part of the 20th century. It took eight years to perform the numerous calculations required. These ephemerides were used unchanged until the beginning of the space era. In 1973, J.H. Lieske took up Sampson's theory and made many improvements, enabling the theory to be adapted to new observations (Lieske 1974). This modified theory, known as the Sampson-Lieske theory, was used by NASA for the Voyager and Galileo missions (Lieske 1996). It was eventually replaced by purely numerical methods on the Galileo mission (Peters 1981).

✉ M. Zannoni
m.zannoni@unibo.it

1  IMCCE, Observatoire de Paris, PSL Research University, Sorbonne Université, CNRS, Univ. Lille; 77 avenue Denfert Rochereau, 75014 Paris, France

2  Dipartimento di Ingegneria Industriale, Alma Mater Studiorum - Università di Bologna, Forli, Italy

3  Institut Polytechnique des Sciences Avancées IPSA, 63 bis Boulevard de Brandebourg, 94200 Ivry-sur-Seine, France

4  Observatoire de la Côte d'Azur, UniCA, Laboratoire Lagrange CNRS UMR 7293, Bd de l'Observatoire, 06304 Nice, France

The study of the orbital motion of the Saturnian moons followed a somewhat similar treatment. The presence of three independent mean motion resonances within the system (Mimas-Tethys, Enceladus-Dione and Titan-Hyperion) led to the development of dynamical models specific to different satellites subsets (Kozai 1957; Sinclair 1977; Harper et al. 1988; Harper and Taylor 1993). A global analytical treatment was eventually given by Duriez and Vienne (1991), Vienne and Duriez (1991, 1992). Analytical developments were then superseded by numerical integration (Jacobson 2004). In the case of the Uranian and Neptunian system, the absence of mean motion resonances made the use of analytical theories more efficient (Laskar 1986; Emelyanov and Samorodov 2015), although numerical integration is mostly used today (Jacobson et al. 1991; Lainey 2008; Emelyanov and Nikonchuk 2013; Jacobson 2014).

Quantifying tidal effects in systems of giant planets from astrometric observations is nothing new. It began in the early 20th century with the work of de Sitter (1928). He targeted the four Galilean moons of Jupiter, which had been observed continuously since their discovery by Galileo in 1610. To do so, de Sitter benefited from a number of observations, many of which he had reduced himself (de Sitter 1916). His early work showed notable accelerations, which we now know to be erroneous. de Sitter's theory was too succinct, given the complexity of the dynamic effects to be modelled. In this respect, the Laplace resonance linking energy and angular momentum of the three moons Io, Europa, and Ganymede will remain a major difficulty in the development of any analytical theory of these satellites. The search for secular tidal accelerations in the moons' motion resumed in the 1970s with the work of Lieske. With Sampson-Lieske's theory in the public domain, researchers began to make their own, sometimes contradictory, measurements (Lieske 1987; Vasundhara et al. 1996). In 2003, Lainey et al. (2004) and Avdyushev (2004) independently found that the Sampson-Lieske theory suffered from a lack of long-period terms, which could often be confused with secular accelerations. Hence, depending on the observations used, former researchers would fit different tidal accelerations. In 2009, the first measurements from purely numerical models appeared (Lainey et al. 2009). The underlying method was already used to numerically quantify the orbital acceleration of Phobos. The main difference is that tidal effects in at least one of the satellites, Io, needed to be introduced. Practically, the dynamical models introduce tides by mean of the ratio $k_2/Q$, where $k_2$ is the degree-2 Love number and $Q$ is the quality factor characterizing the dissipation of tidal energy in the planet.

The extension of this work to the Saturn system was quite natural in the context of NASA's Cassini mission. Although initial estimates of secular accelerations had been found by Kozai (1957) and Dourneau (1987), it would appear that, as with the Jupiter system, it was the absence of a long period in the analytical theory that was lacking (Vienne et al. 1992). The first measurements from a numerical model were published in 2012 by Lainey et al. This work was then enhanced by the introduction of the numerous measurements made by the ISS/Cassini instrument (Porco et al. 2004) to produce the first measurement of Saturn's $k_2$ (Lainey et al. 2017), as well as the frequency dependence of the dissipative factor Q. Last but not least, radio science data revealed to be of high importance for constraining the tidal expansion of Titan's orbit (Lainey et al. 2020).

In Sect. 2 we present the available astrometric methods that have been used to determine tidal parameters in giant planets. Section 3 focuses on radio science data. The next section details the various measurements made for the Jupiter system. Section 5 presents the latest results for the Saturn system. In Sect. 6, we discuss the first measurements for the Uranus system and possible future ones for Neptune system. Section 7 focuses on a few extra-solar systems.

## 2 Astrometry

### 2.1 Classical Astrometry

Astrometry – referring to positioning astronomy – is the first science deriving from observational astronomy. It allows the determination of spherical coordinates in right ascension and declination of the observed objects, to assess their motion and dynamics, and to determine parameters of their physical environment (masses, internal mass distribution, tidal parameters...), in particular. As the methods of observation and data acquisition have benefited from technological advances with the transition from analog (astrophotographic plates) to digital (CCD and CMOS sensors), measurement methods have also evolved with the development of new techniques, which now complement the classical approach of plate constants calibration. Finally, these data are used in conjunction with a dynamic model to quantify physical parameters, including tides.

#### 2.1.1 Photographic Plates

The astrophotographic plates were to the telescope what the photographic films were to the camera: analog supports to acquire and save shots. They were produced by the manufacturers of the epoch (Kodak, Agfa-Gevaert...), used since the end of the 19th century to take pictures of scenes of life and by extension, astronomical observations (Fig. 1). They have been progressively replaced by numerical CCD sensors since the end of the 1990's, and the production has stopped.

Photographic plate formats could be very different depending on their use for astrometry and photometry (example of a 35x35 cm Schmidt plate), or spectroscopy (example of a 2x25 cm star spectra). One side was covered by a photosensitive emulsion whose silver halides blackened under a light beam. This was followed by the classical photographic development phases in a darkroom to fix the emulsion and clean the support. Exposure times range from several seconds to a few hours depending on the observed target. The size of the field, depending on the instrumentation and the size of the plates, varies from a few arcminutes to several degrees on the celestial sphere.

Due to their analog nature, it was possible to filter the plates by direct apposition, partial or complete, with the huge advantage to allow for simultaneous observation objects with very different magnitudes. Another particularity, due to the relatively high cost of this equipment: it was very common to carry out several exposures on the same support by shifting the plate in right ascension or declination according to the mode of mounting at the focus of the telescope, between exposures.

Firstly analyzed by hand, photographic plates could now be digitized using specific machines to provide numerical images to be re-studied with new numerical tools, providing new accurate information of past observations (Robert et al. 2021). Moreover, because relatively few photographic plates have already been analyzed, mainly for the needs of the first space missions (Mariner, Viking...), we there find a huge observation tank with original data allowing the adjustment periods to be extended in dynamical models.

#### 2.1.2 CCDs

Charge-Coupled Devices, CCDs are digital photographic sensors developed since the 1970's, fully by the end of the 20th, and now commonly used for industry, scientific imagery, and general public (smartphones, reflex cameras...).

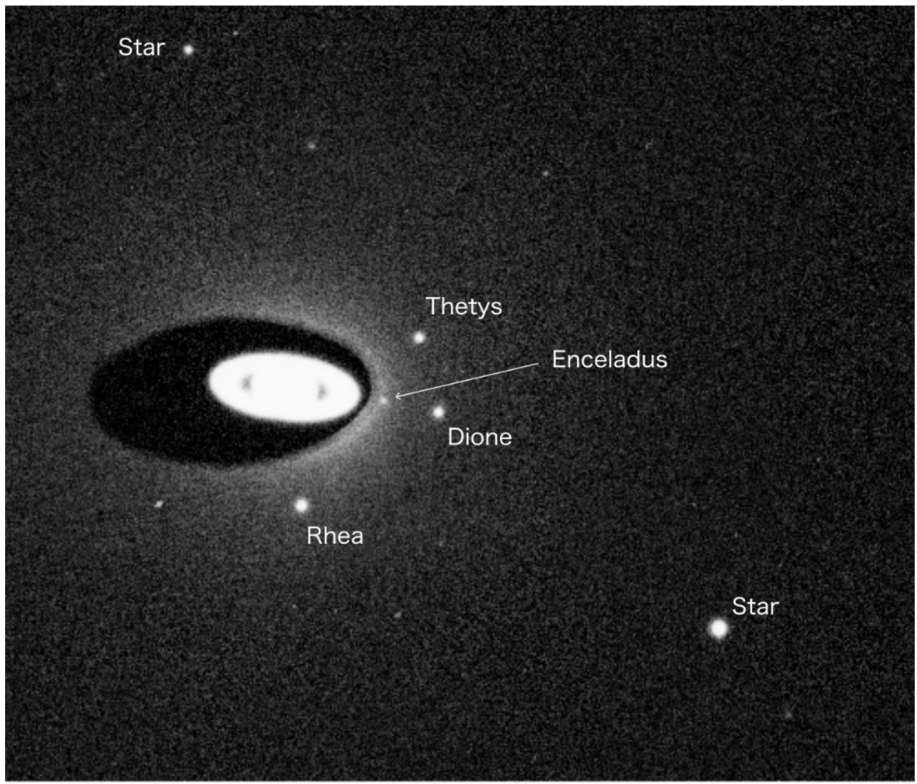

**Fig. 1** Example of a USNO photographic plate (center part) of Saturn in 1974, showing stars and four of the main satellites

CCDs are generally composed with semiconductor elements able to convert photons into electrons that are stored in a memory cell before decoding. They are mounted as pixel lines or pixel matrices. They are characterized by their quantum efficiency (the ratio between the number of incoming photons and the number of generated electrons), their transmission (wavelength response), resolution, pixel size and number, information encoding... Depending on their architecture, they could generate monochrome or RGB colored images.

CCD cameras provide numerical images that can be used in astrometry, photometry, and spectroscopy (Fig. 2). They directly benefit from numerical tools for image transfer, treatment, analysis and archiving. Apart from their intrinsic analogic-numeric difference, CCDs have a quantum efficiency that can peak at 90% compared to 2-10% for photographic plates (Howell 2006), arguing in favour of the replacement of these former materials. Moreover, CCDs allow to observe objects by considerably decreasing the exposure time, without sensibility loss. When observing natural satellites, exposure times range typically from some tenths of second to several minutes depending on the observed target. The size of the field, depending on the instrumentation and the size of the sensor is up to some tens of arcminutes on the celestial sphere.

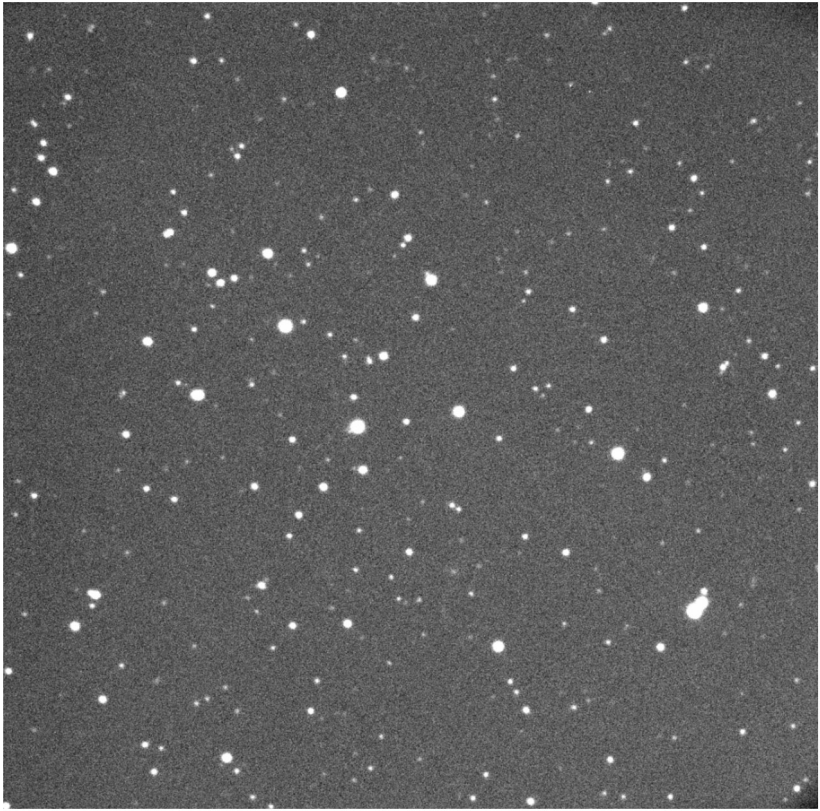

**Fig. 2** Example of a monochrome OHP-T120 CCD observation of Pluto in 2023, July

### 2.1.3 Precision Premium

Precision premium is a complement to the classical astrometric techniques, first presented by Pascu (1994) within an application to the Jovian system. It consists of observing a pair of close moons using the other moons present all around in the field for calibration. Indeed, a major source of error often comes from an inaccurate image scale factor. Since relative positions on the tangential plane depend essentially linearly on the scale factor, the use of close pair of moons is extremely beneficial.

Historically, before the arrival of the first accurate astrometric star catalogs, astronomers could not link the reference stars on the celestial sphere to calculate the observation plate constants, which are essential to determine spherical coordinates in right ascension and declination. Thus, only relative (intersatellite) positions could be used for the adjustments of dynamical models. This has the side advantage to provide measurements independent on the planetary ephemeris error.

Precision Premium is an observation concept that constraints the realization of the observations to provide the most precise astrometric results. One could also consider Precision Premium as a condition of observation sampling since it can refer to very specific parts in large data set. Within his application, Pascu (1994) wrote that the most accurate relative positions of two objects should be obtained when their apparent angular separation is small. It was demonstrated with observations of the Galileans with separations below 50 arcseconds. More recently, Lin et al. (2019) showed that the range of separation for which Precision Premium could be used is up to 100 arcseconds, allowing to decrease the object positioning errors up to several milliarcseconds.

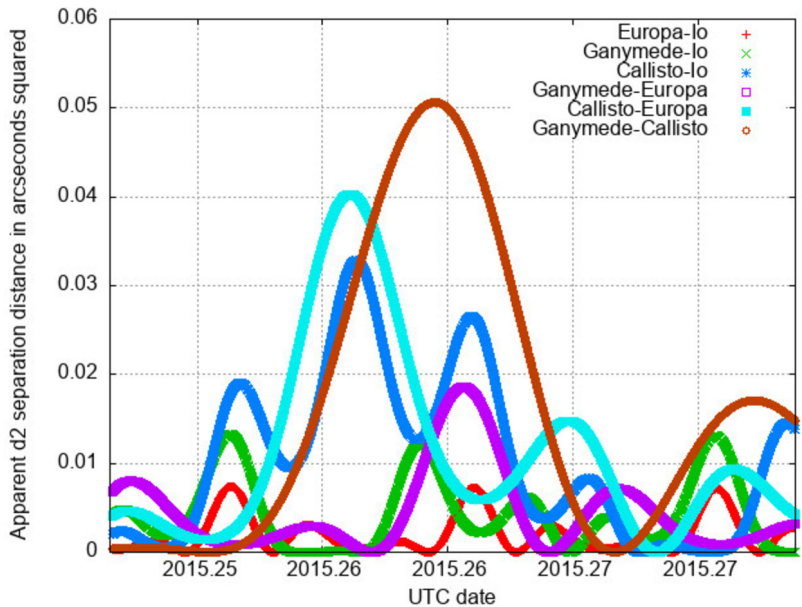

**Fig. 3** Occurrences of minimum and maximum intersatellite distances over 10 days

Precision Premium has been used for intersatellite measurements, but it can also be generalized to all inter-object positioning with the same accuracy. Moreover, it allows the decorrelation of the motion of the measured bodies even if they are rotating around a primary.

### 2.1.4 Mutual Approximations

Mutual approximations (Morgado et al. 2016) were somewhat inspired from the mutual events observation (e.g. mutual occultation or eclipses of one moon by another). The apparent relative motion of satellites allows the observation of the approaches and distances. We no longer fit magnitude drops but distance drops. Thus, minimum and maximum distance times can be used to constrain dynamical models. The geometric parameters of central instant, impact parameter and relative velocity that describe mutual events are the same for mutual approximations.

Mutual approximations occur several times each day, and can be continuously observed to provide regular accurate positioning (Fig. 3).

### 2.2 Astrometry Derived from Photometry

Although not strictly speaking an astrometric measurement, the observation of celestial phenomena involving moons is often one of the most reliable means of doing an astrometric measurement. The longest observed phenomena are eclipses of moons, particularly Galilean moons. For a long time, these have served as a reliable time reference for deducing longitude measurements of places, thus improving our knowledge of global cartography (Lieske 1986a). These easy-to-make measurements continued to be used until the beginning of the 21st century (Mallama et al. 2010). However, the observation of so-called mutual phenomena, involving the eclipse or occultation of one moon by another, made it possible in the 1970s to considerably increase the precision of astrometric measurements of the moons of Jupiter (first campaign in 1973) and Saturn (first campaign in 1979).

The clear advantage of observing phenomena is that they are not an angular measurement, but a photometric one. This means that, provided the signal-to-noise ratio is sufficiently high, the astrometric measurement (deduced from the photometric light curve) is much less dependent on the distance of the observed system than conventional astrometric measurements, for which a precision of a tenth of an arcsecond will not give the same kilometric precision within the system, depending on whether it is near or far from the Earth. In this respect, the campaign to observe the mutual phenomena of Uranus' satellites enabled us to obtain measurements of a quality unattainable from the ground using conventional methods (Arlot et al. 2013). Today, the processing of observations of mutual phenomena is limited by two main factors. The first is the lack of albedo maps of objects at the observed wavelengths (Vasundhara et al. 2003; Emelyanov 2017). The second is the difficulty often encountered in obtaining detailed information on the observing conditions of many observers like detector characteristics, bandwidth of filters used, etc.

However, even the observation of mutual phenomena has now been superseded by an even more powerful technique: the observation of star occultations. While this technique is by no means new in itself, it is the presence of the Gaia stellar catalog that now enables this observational technique to take the lead in moon astrometry (Morgado et al. 2019). Thanks to observations made at various locations on Earth, it is possible not only to determine the shape of the occulting object, but also its precise position on the celestial sphere. Accuracies of just a few milliarcseconds can easily be achieved, which is an order of magnitude more precise than the observation of mutual phenomena.

### 2.3 Space Astrometry

Astrometric processing of space images is not fundamentally different from that used for ground-based images (see Sect. 2.1). When the conditions are right, we can generally find enough stars in the image background to perform a field calibration. The presence of star trackers on probes often means that only a minor correction has to be made to the center of the field, although this is absolutely essential for a quality measurement. The camera's optical parameters are generally obtained by pointing at star clusters (Owen 2003). The essential differences therefore concern the treatment of star photocentres (Zhang et al. 2021) and the determination of the geometric center of resolved celestial objects (Tajeddine et al. 2013).

The Caviar software (Cooper et al. 2018), available under an open source license, enables precisely this type of measurement (Fig. 4). It was used throughout the Cassini mission to provide astrometric data for the moons of the Saturn system. A typical accuracy of a few tenths of a pixel is frequently obtained, which amounts to an accuracy of between a hundred meters and a few kilometers, depending on the probe's distance from its target. It should be noted that the accuracy of these measurements is sometimes worse than the precision. Among the sources of error that may explain these measurement biases, the difficulty of precisely defining the position of the edge of objects has been reported (Cooper et al. 2014), as has the relative lack of precision of the ellipsoidal models used for images taken at close range to the target (Cooper et al. 2018).

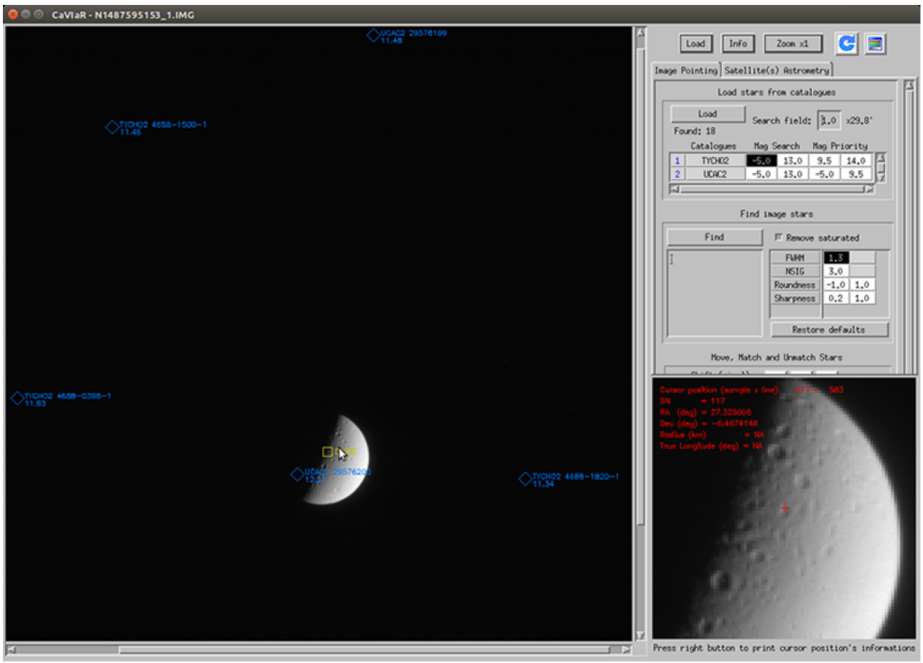

**Fig. 4** Screenshot of Caviar software available under open source license (Cooper et al. 2018). Dione can be seen in the image used N1487595153_1 (available on https://pds.nasa.gov/)

## 2.4 Gaia

Although Gaia is a space mission, the astrometric processing carried out is somewhat different from the standard procedure. In fact, the Gaia probe scans the sky according to the evolution of its axis of rotation. Gaia's astrometric measurements are so precise because they are based on a measurement of time, not space. As the probe rotates, the stars move across the focal plane. Since Gaia's rotation is perfectly well known, measuring the time taken for the various objects to pass over the nine CCDs enables to obtain an extremely precise relative position of the stars in relation to each other in the direction of Gaia's movement. Measurements of a few tenths of a milliarcsecond are common for moons in the solar system (Tanga et al. 2023). Measurements are therefore essentially made in a single direction, although a measurement perpendicular to the axis of rotation is also made, albeit with much lower precision (typically 0.6 arcseconds).

A number of technical difficulties somewhat limit Gaia's contribution to the solar system's moons motion. Firstly, moons that are too close to their planet are not detectable. Secondly, measurement of very large objects such as Jupiter's four Galilean satellites or Titan are not reliable. Last, objects of magnitude below 20.9 are observed by Gaia, only. Nonetheless, for outer moons and a good number of main moons, Gaia data are exceptionally accurate. So, while the first measurements of the solar system's moons have not been provided until DR3 (Gaia Collaboration 2023), it's pretty clear that Gaia astrometric data is becoming a remarkable asset for the possible quantification of tidal dissipation in Uranus and Neptune, especially after the publication of Gaia DR5 release that will provide 10 years of high accuracy astrometric measurements.

### 2.5 Radar Measurements

Radar measurements are another way of obtaining direct astrometric information about the position of moons. The principle is to measure the round trip made by a series of radar shots between the transmitting antenna and one of the moons. This can be done from an antenna on the ground or from a space probe. In the first case, the great distance separating the Earth from the giant planets makes it necessary to use a very large antenna. Until now, it has only been possible to make observations of the Galilean satellites from the Arecibo radar station (Harmon et al. 1994; Brozovic et al. 2020). The first measurements were made of the moons Ganymede and Callisto in 1992, then Europa in 1999 and finally Io in 2015. A set of 22 measurements taken over the period 1992-2016 is now available. The precision obtained ranges from 1.5 km for the best observations to several tens of kilometres for the least precise (Brozovic et al. 2020).

More recently, radar shots taken by the MARSIS instrument on board the European Mars Express probe have also been analysed for astrometric purposes. A measurement accuracy of a few hundred metres was obtained for the position of Phobos (Desage et al. 2024). It is therefore likely that similar measurements will be possible with the arrival of the Juice probe in the Jovian system in 2031 (Van Hoolst et al. 2024).

### 2.6 Orbital Dynamics

All these astrometric data can be compared with the predictions of orbital models for the moons concerned. These models, formerly analytical and now numerical, consist of an integration of the equations of motion. Starting from approximate initial conditions, it is then possible to study the discrepancy between observed and calculated positions. From these deviations, a correction can be made to the physical parameters modelled within the dynamical system, as well as to the initial conditions. To do this, one also needs to know the derivative of the position and velocity vectors with respect to these quantities at the observation dates. In practice, it is often necessary to integrate an additional system of differential equations, also known as variational equations. Ultimately, the inversion of a linear system is performed using least-squares techniques. More details on these aspects can be found in Lainey (2016).

## 3 Radio Science

Radio science measurements are fundamentally different from astrometric ones because they do not refer directly to the target bodies and, therefore, they have a different information content. The main radio science observables are two-way range and range-rate (Doppler) measurements between the spacecraft and the ground station's antennas, enabled by an onboard transponder (Fig. 5). Range-rate measurements are obtained from the Doppler shift of a highly stable microwave carrier, while range is obtained from the round-trip light-time of a ranging code modulated onto the radio signal. The radio link provides direct information on the state of the spacecraft. However, since its trajectory is affected by the gravity of close celestial bodies, it is possible to derive the 3-d position of those bodies through an orbit determination process. The core of this process is the comparison between the measurements obtained from the spacecraft and the ones computed by means of mathematical models (Milani and Gronchi 2009). To better fit the observations, the trajectory of the spacecraft and the orbits of close celestial bodies must be precisely reconstructed, so a set of parameters that affect their dynamics must be jointly estimated. The dynamical model must include all the non-negligible forces that, for missions at giant planets, typically include: the point-mass gravity of the Sun, the planets of the Solar System, and close moons, including relativistic effects; the extended gravity of the close celestial bodies, usually in terms of spherical harmonics coefficients; rotational models of the celestial bodies that have an extended gravity field; time-varying gravity due to tidal interactions; non-gravitational forces, such as the solar radiation pressure and atmospheric drag (e.g. Lainey et al. 2020; Gomez Casajus et al. 2022).

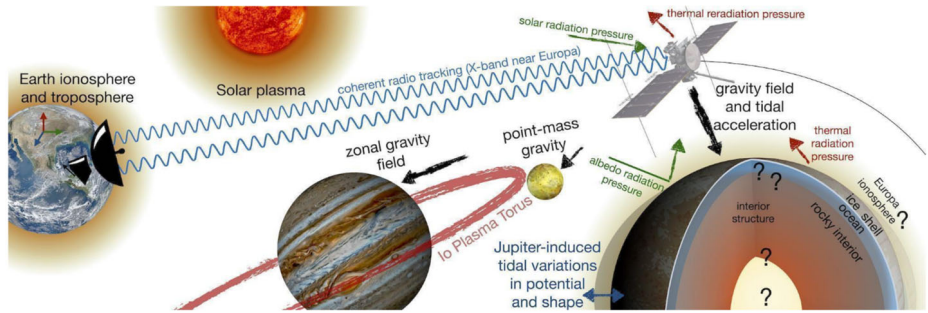

**Fig. 5** Illustration of radio science investigations of a deep space mission (from Mazarico et al. 2023). The main radio science measurements are derived from the coherent radio link with Earth ground station antennas. They provide information on the spacecraft trajectory, influenced by both gravitational and non-gravitational accelerations

The quality of radiometric measurements depends on many elements of both the space and ground segments, but the main driving factor is the carrier frequency adopted in the radio link. Higher frequencies are less affected by dispersive media, among which the solar plasma is typically the dominant source of noise in the data (Asmar et al. 2005; Iess et al. 2014). With time, space exploration missions adopted higher and higher carrier frequencies. The S-band link (2.1-2.3 GHz), used by the Pioneer, Voyager, and Galileo probes, allowed accuracies of the order of 1 mm/s (Doppler, 60 s integration time) and 10 m (ranging). Moving to X-band (7.1-8.4 GHz), used for example by the standard Cassini radio link, improved the accuracies to about 0.05 mm/s (Doppler, 60 s integration time) and 2 m (ranging). Finally, the Ka-band link (32-34 GHz), adopted by Juno (Doppler only), BepiColombo (Doppler and ranging), and Juice (Doppler and ranging), allows reaching accuracies of better than 0.03 mm/s (Doppler, 60 s integration time) and 1 m (ranging). In addition, the availability of multiple radio links at different frequencies, guaranteed by state-of-the-art radio science instruments such as the ones carried by Cassini (up to 2003), Juno, BepiColombo, and Juice, allows the removal of almost completely the dispersive noises (Bertotti et al. 1993; Mariotti and Tortora 2013), leaving the Earth troposphere as the residual leading noise source. State-of-the-art tropospheric calibrations are produced through water-vapour radiometers, currently available only at NASA's DSS 25 and ESA's Malargue stations (Buccino et al. 2021; Lasagni Manghi et al. 2021). When multi-frequency links and water-vapour radiometers are combined, accuracies of the order of 0.01 mm/s (Doppler, 60 s integration time) and 20 cm (ranging) can be reached, as demonstrated by Juno (Durante et al. 2020) and Bepi-Colombo (Lasagni Manghi et al. 2023), opening the way to next-generation radio science investigations (Iess et al. 2021, 2024).

When a spacecraft is in orbit around a body, range-rate measurements allow to reconstruct the relative orbit of the spacecraft with respect to that body. Complementary range measurements provide strong information on the body's orbit, making them the main observables for the development of the planetary ephemerides (Folkner et al. 2014; Park et al. 2021).

When the spacecraft performs a flyby of a celestial body, most of the information content comes from range-rate measurements. During each encounter, these measurements provide the relative position of the spacecraft with respect to the body. Before and after the encounter, when the spacecraft is outside the sphere of influence of the body, range-rate measurements provide the relative position of the spacecraft with respect to the central body (the planet, in case of moon flybys). As a result, precisely reconstructing the trajectory during the entire encounter allows us to get the relative position of the body with respect to the central body. Typical X-band accuracies in the retrieved position of moons with a flyby are in the order of few hundreds of meters on the orbital plane, tens of km normal to the orbital plane. However, because flybys provide high-accuracy but rather sparse data, to correctly reconstruct the satellite ephemerides radio science data must be properly combined with measurements providing better coverage, such as astrometry.

### 3.1 Voyager

In 1979, Pioneer 11 became the first space mission to collect radio science data of the natural satellites of Saturn, allowing to estimate the gravity field of Saturn, together with the masses of Rhea, Titan, and Iapetus (Null et al. 1981). However, the satellite ephemerides were taken directly from analytical theories based on astrometric measurements. Later, in 1980 and 1981 the twin Voyager probes performed a flyby of Saturn. A joint analysis of Pioneer and Voyager radio science data, together with Voyager astrometry allowed to improve the estimation of the gravity of the Saturn system (Campbell and Anderson 1989). In addition, parameters of the analytical theory describing the orbits of Tethys, Rhea, and Titan were estimated.

Later, in preparation for the Cassini mission, the addition of Earth-based astrometry to radio science and space-based astrometry allowed an improved and consistent estimate of the gravity field of Saturn, and the masses and orbits of the Saturnian moons, through numerical integration of a full dynamical model (Jacobson 2004). The resulting uncertainty in the moons' positions during the analysed timespan was ∼50 km in the radial and normal directions and 150-500 km in the transverse direction, not sufficient to detect tidal dissipation effects in Saturn.

### 3.2 Cassini

The Voyager probes returned a huge impact in science and helped to understand better the satellite systems of the gas giants. However, they left many unanswered questions about Saturn and its moons until 2004, when the Cassini spacecraft entered into orbit around the planet. During its 13-year mission, it performed many close flybys of the moons of Saturn, including 127 flybys of Titan. Radio science data were collected during the closest approach of 3 flybys of Enceladus, 2 of Rhea, 3 of Dione, and 10 of Titan. During the mission, the satellite ephemerides of the Saturn systems were periodically updated by the Cassini team to support the navigation and science activities. These analyses made use of an extensive data set comprising all available radio tracking data of Cassini, Pioneer 11, and Voyager, space-based and Earth-based astrometric data, including stellar and ring occultations (Jacobson et al. 2006; Jacobson 2022).

In 2017, the Cassini mission ended with a "Grand Finale", a series of inclined, highly eccentric orbits characterized by a very low pericenter, with altitudes 2600-3900 km above the cloud tops. Radio science data were collected during five of these pericenter arcs, allowing to estimate the gravity field of Saturn at an unprecedented level of accuracy, including for the first time the mass of its rings (Iess et al. 2019). A time-variable component of the Saturn gravity field was also detected, possibly related to the planet's normal modes (Markham et al. 2020), similar to what was later observed in Jupiter by the Juno mission (Durante et al. 2022).

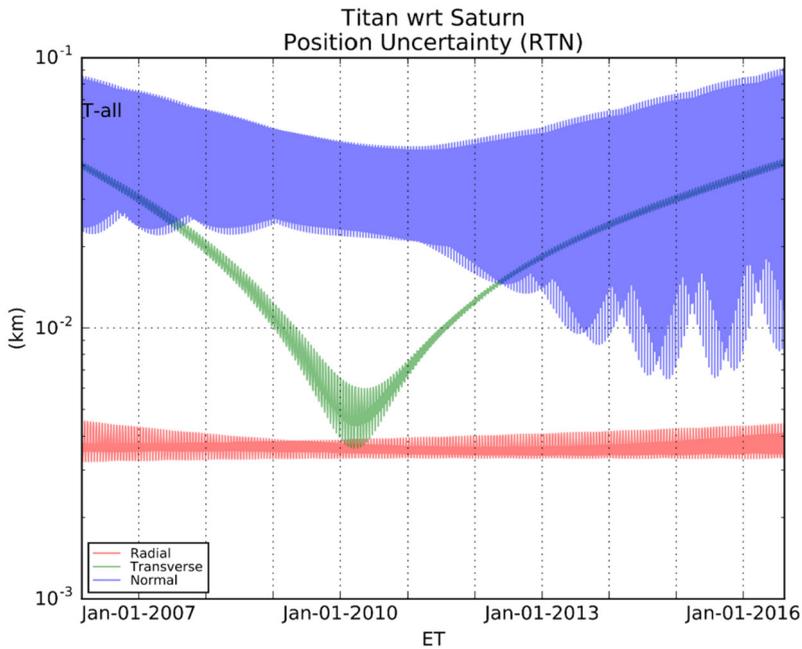

**Fig. 6** Uncertainties of Titan trajectory with respect to Saturn in the radial, transverse and normal to the orbital plane directions, obtained from Cassini radio science data (from Gomez Casajus 2019)

The gravity field of Saturn observed during the Grand Finale also allowed to better characterize the orbital motion of the main Saturnian satellites. In particular, given the overall good quality of radio tracking data and the good time coverage of Titan flybys, Lainey et al. (2020) precisely reconstructed a coherent orbit of Titan for the entire timespan of the Cassini mission, using only radiometric data (see Sect. 5.2). The obtained formal uncertainties in the Titan's position are 3 m in the radial direction, almost constant during the timespan of the Cassini mission. In the transverse direction, the uncertainty increases almost linearly from a minimum of 4 m in January 2010 to a maximum of 40 m. The sensitivity in the orbit normal direction is lower, with uncertainties up to 50-100 m, during the entire mission (Fig. 6).

### 3.3 Galileo

The Galileo mission studied the Jovian system from December 1995 to September 2003, dramatically increasing our knowledge of the system, and especially of its Galilean moons, thanks to more than 30 close flybys. Radio science data were acquired during most of these close encounters. Unfortunately, the overall scientific return of the mission was significantly reduced by the missing deployment of the umbrella-like high-gain antenna, forcing to communicate with the spacecraft using only the low-gain antenna. The quality of radio science data was heavily affected because the low-gain antenna supported only a low signal-to-noise ratio radio link at S-band, resulting in high noise levels due to solar plasma and station electronics. In addition, the unavailability of a multi-frequency link at both S- and X- bands, as originally envisioned by the high-gain antenna radio link, prevented calibration of the Doppler shift caused by the ionospheres of Jupiter, its moons, and the Io plasma torus, possibly introducing biases in the orbit determination results.

As a result, using only Galileo radio tracking data acquired during the moons' close encounters, the obtainable formal uncertainty in the moons' positions with respect to the central body was a few kilometers in the radial and transverse directions and tens of kilometers in the orbit normal direction.

Nevertheless, during and after the Galileo mission, the orbits of the Galilean moons, together with their gravity and the gravity of Jupiter, were periodically updated jointly analysing Galileo and Ulysses radiometric data, space-based and Earth-based astrometry, including mutual events and eclipses (Jacobson et al. 2000; Jacobson 2001).

### 3.4 Juno

The Juno mission, dedicated to the study of the origin and evolution of Jupiter, reached the planet in July 2016, when it entered into a highly eccentric, polar orbit characterized by very low pericenter altitudes (less than 5000 km above the cloud tops). The orbit was specifically designed to avoid Jupiter's intense radiation belts and, at the same time, to provide a detailed characterization of the planet. In particular, Juno measured the gravity field of Jupiter at an unprecedented level of detail (Durante et al. 2020), including the possible detection of normal modes (Durante et al. 2022).

Because of Jupiter's oblateness perturbation, Juno's orbit periapsis precesses northward, allowing the inclusion of 4 close encounters of the inner Galilean satellites after the end of the prime mission in 2021: one of Ganymede (June 2021), one of Europa (September 2022) and two of Io (December 2023 and February 2024). During the flybys, radio tracking measurements have been acquired, offering the opportunity to update the moons' gravity and orbits for the first time since the end of the Galileo mission (Gomez Casajus et al. 2022). This opportunity was particularly interesting for Io, which will not be visited by any of the currently approved missions to the Jupiter system, namely Juice and Europa Clipper (Park et al. 2025).

Moreover, as opposed to Galileo, Juno is equipped with a state-of-the-art radio system that supports the acquisition of Doppler measurements at both X and Ka-band, allowing to reach noise levels of 0.01-0.03 mm/s (60 s integration time) and to remove the Doppler signatures caused by the ionospheres. However, Juno's orbit is not optimized for moon investigations, and the very high velocities relative to the moons decrease the sensitivity of radio science measurements. As a result, the formal uncertainty in the Ganymede's position obtained during the G34 close flyby was 100 m in the radial direction, 200 m in the transverse direction, and 5 km in the orbit normal direction (Gomez Casajus et al. 2022). The Juno radio science measurements of the moons, jointly analysed with Galileo data, astrometric data, together with the radio data of the future Juice and Europa Clipper missions, have the potential to significantly improve our knowledge of the orbit of the moons and the tidal dissipation of the Jupiter system (Magnanini et al. 2024; Fayolle et al. 2023).

**Table 1** Tidal acceleration under the form of $\dot{n}/n (10^{-10} \ yr^{-1})$ where n stands for the mean motion and determined from observations. Even though Io is expected to suffer tides more than Europa and Ganymede, the presence of the Laplace resonance dispatches the change of Io's orbital energy into the three innermost Galilean moons. It is noteworthy that Lieske (1987), Lieske (1986a) and Aksnes and Franklin (2001) used the Sampson-Lieske theory to fit their data

|  | Io | Europa | Ganymede |
| --- | --- | --- | --- |
| de Sitter (1928) | +3.3 +/− 0.5 | +2.7 +/− 0.7 | +1.5 +/− 0.6 |
| Lieske (1987) | −0.074 +/− 0.087 | −0.082 +/− 0.097 | −0.098 +/− 0.153 |
| Goldstein and Jacobs (1995) | +4.54 +/− 0.95 | +5.6 +/− 5.7 | +2.8 +/− 2.0 |
| Lieske (1986a) | +2.27 +/− 0.70 | −0.67 +/− 0.80 | +1.06 +/− 1.00 |
| Aksnes and Franklin (2001) | +3.6 +/− 1.0 | - | - |
| Lainey et al. (2009) | +0.14 +/− 0.01 | -0.43 +/− 0.10 | -1.57 +/− 0.27 |

## 4 The Jovian System

### 4.1 The Early Measurements

De Sitter was probably the first to measure the expected tidal acceleration of the Galilean moons. Benefiting from heliometric instruments and photographic plates, he was able to envisage the detection of tidal accelerations. Unfortunately, the complexity of the dynamics involved in the large masses of the Galilean moons made it extremely difficult to develop an analytical solution for orbital motions. This question remained topical throughout the twentieth century, notably with the famous work of J.H. Lieske. Lieske put a great deal of effort into demonstrating the expected tidal accelerations. Interestingly, he collected 16,842 of eclipses data, some dating back to the mid-17th century (Lieske 1986a,b). If the connection of the time scale used at the time with the modern time scale was a problem, so too was the absence of long periodic terms in the so-called Sampson-Lieske theory.

It was only at the beginning of the 21st century that the dynamic inconsistencies of this orbital theory came to light (Lainey et al. 2004; Avdyushev 2004). In particular, the absence of long periodic terms could easily be confused with tidal accelerations. As many observers used Sampson-Lieske theory to adjust tidal accelerations, they obtained inaccurate results. Interestingly, depending on the period covered by the astrometric data used by the observers, their results simply seemed to contradict each other (Table 1).

### 4.2 Using Numerical Models

Benefiting from numerical simulations, the first accurate estimation of tides inside Jupiter and Io was given by Lainey et al. (2009). Introducing the Jovian $k_2$ from Gavrilov and Zharkov (1977) and assuming a constant quality factor Q for Jupiter, they found that Io's orbit is decaying, while Ganymede's and Europa's orbits are expanding. The fitted dissipation values correspond to orbital acceleration values (dn/dt)/n of +0.14+/−0.01 x $10^{-10}$ yr$^{-1}$; −(0.43+/−0.10) x $10^{-10}$ yr$^{-1}$; and −(1.57+/−0.27) x $10^{-10}$ yr$^{-1}$; (formal error bars, 1-sigma) for Io, Europa and Ganymede, respectively. These accelerations represent a shift in the satellite orbital positions of respectively 55 km, 2125 km and 2365 km over the 116 years considered. Using such numerical values, the study of the Laplace resonance angle suggests that the Laplace resonance is evolving, which could be the consequence of tidal cycles within Io (Ojakangas and Stevenson 1986). More recently, Park et al. (2025) re-estimated the tidal dissipation within Io and Jupiter. They found a solution consistent with Lainey et al. (2009). However, in the continuation of former works, such results assumed a constant Jovian Q value for all tidal frequencies. How far such an approximation is reliable remains an open question. In particular, computation of tidal dissipation inside planets with a fluid envelope and solid core predicts a strongly frequency dependent dissipation (e.g. Ogilvie and Lin 2004). This suggests that the error bar on the Jupiter and Io tidal ratio $k_2/Q$ could be underestimated.

### 4.3 Love Numbers Estimation

The knowledge of the gravity field of Jupiter was revolutionized by the Juno spacecraft, which has been in orbit around the gas giant since 2016. Its polar orbit, characterized by a very low pericenter radius, allows a strong sensitivity to many of the components in which the planet's gravity can be decomposed. Given an isolated non-rotating body in space, its internal mass distribution is expected to be spherically symmetric, and the corresponding external gravity field is equivalent to that of a simple point mass. Because of the rotation around its spin axis, the shape and mass distribution change under the combined effects of internal pressure and the centrifugal force. If the body is in hydrostatic equilibrium, which is a good assumption over long time scales, especially for a gaseous planet, the shape of the planet will follow an equipotential surface. For a gaseous planet rotating rigidly around its principal polar axis of inertia, the mass distribution is expected to be symmetric around the spin axis and between the North and South hemispheres. The resulting gravity field can be represented only by zonal harmonics (order m=0) of even degree (l=2n), whose magnitude decreases with the degree (solid body rotation). Thanks to the favourable orbit and the state-of-the-art radio science instrumentation, Juno was able to measure the even zonal harmonics coefficients up to degree 10 (Serra et al. 2019; Durante et al. 2020), allowing to characterize its interior structure to an unprecedented level of detail (e.g. Stevenson 2020; Helled et al. 2022; Miguel et al. 2022; Militzer et al. 2022).

In addition to the dominating solid body rotation, many smaller effects perturbing the mass distribution symmetry are at play. First, the dissipation of internal energy sustains the atmospheric dynamics visible on the upper layer of the gas giants, causing variations in the interior pressure and density and, thus, in the gravity field. The largest effect is represented by the zonal winds, which exhibits a symmetry around the spin axis but are asymmetric between the North and South hemispheres. As a result, the gravity field also presents smaller odd zonal harmonics. Juno detected for the first time the odd zonal harmonics coefficients up to degree 9 (Iess et al. 2018), helping constraining the depth of the zonal winds to about 3000 km (Kaspi et al. 2018), after which the planet rotates nearly as a rigid body (Guillot et al. 2018). In addition to the zonal winds, smaller-scale vortices introduce non-zonal, time-varying terms in the gravity field expansion, whose amplitude is expected to be smaller than the Juno sensitivity. However, Juno data allowed to detect the gravity signature of the great red spot, the biggest atmospheric vortex in Jupiter's atmosphere, constraining its depth to less than 500 km (Parisi et al. 2021).

Besides atmosphere dynamics, normal modes of giant planets are, in principle, able to perturb the interior density profile, producing a time-varying gravity field. Accumulating the data of many perijove passes, Juno experienced a time-varying gravity field that could be explained by normal modes, in particular by p-modes with a peak radial velocity at the surface of 10-50 cm/s at 900-1200 $\mu$Hz (Durante et al. 2022).

Finally, Jupiter's shape and distribution of mass are perturbed by the tidal potential generated by its natural satellites, among which the dominant effect is due to the Galilean satellites and especially Io, because of its combination of mass and distance. Each satellite produces a tidal bulge on the planet, oriented mainly along the direction of the satellite itself, where a small misalignment is expected due to the planet's internal dissipation, equivalent to time and phase lags. The resulting effect on the planet's gravity field can be modeled through a time variation of the spherical harmonics coefficients, parameterized by a complex Love number $k_{l,m}$ for each degree l and order m, where the imaginary part is related to the tidal dissipation. Up to its mid-mission in December 2018, Juno was able to detect the variation of the gravity field due to the different orientations of the tidal bulge between the perijove passes, providing a statistically significant measure of the real part of the Love numbers $k_{2,2} = 0.565 +/- 0.018$, $k_{3,1} = 0.248 +/- 0.046$, $k_{3,3} = 0.34 +/- 0.12$, $k_{4,2} = 1.29 +/- 0.19$, and marginally $k_{4,4} = 0.55 +/- 0.41$, assumed constant for all satellites (Durante et al. 2020). In particular, the Love numbers $k_{2,2}$ and $k_{4,2}$ were found to be different than the values coming from interior model predictions for a static tidal response to Io (0.590 and 1.74, respectively; Wahl et al. 2020), possibly caused by a dynamical response (Idini and Stevenson 2021, 2022a,b; Dewberry and Lai 2022; Dewberry 2023; Lin 2023). However, interior models predict that Jupiter's static tidal response varies with the distance to the perturber because of its large oblateness (Wahl et al. 2020), meaning that a different Love number has to be expected for each satellite. Considering satellite-dependent tides, Juno estimation capability degrades, only the Love numbers at Io's frequency are observable, and $k_{2,2}$ and $k_{4,2}$ become statistically compatible with theory without the need for dynamical tides (Durante et al. 2020). The smaller sensitivity of Juno data to satellite-dependent tides is caused by the fact that Juno's orbital period (∼53 d) is close to resonance with Io and Europa, so the longitudinal separation between the two moons during the perijove passes is almost constant, making difficult to distinguish the different tidal bulges (Notaro et al. 2019). However, during the extended mission, Juno performed several close encounters with the Galilean satellites (see Sect. 3.4), which decreased its orbital period to 43 days after Ganymede's flyby, to 38 days after the one Europa, and finally to 33 days after the two flybys of Io. This would disentangle the tidal effects of Io and Europa, potentially allowing to obtain a reliable frequency-dependent tidal estimation and detect deviations from the static theory.

Finally, Juno does not allow to characterize the tidal dissipation within Jupiter. In fact, the data are directly sensitive to the gravity field variation caused by the tidal bulges, so that a comparable estimation accuracy can be achieved on both the real and imaginary parts of Jupiter's Love numbers. If the obtained accuracy provides a good estimation of the real parts, the imaginary parts are expected to be much smaller in value, preventing a statistically significant estimation.

## 5 The Saturnian System

Although further from Earth than Jupiter, the Saturn system has a number of features that make it ideal for observational determination of tidal effects within this system. First of all, the number of moons is much greater, making it easier to retrieve information and obtain a global solution. But above all, there is no resonance involving at once the majority of the system's moons, as is the case with the Laplace resonance in the Jovian system. Irrespective of the crucial importance of the Cassini mission, these assets have certainly helped a great deal in obtaining the tidal parameters within this system.

### 5.1 Mid-Sized Moon Orbital Expansion

As mentioned in the introduction, Kozai (1957) and Dourneau (1987) determined some first estimations of tidal secular accelerations in the Saturnian system. But similarly to the Galilean system, long periodic terms were missing in their dynamical model making their observation irrelevant to tides (Vienne et al. 1992). The first accurate quantification of Saturn's $k_2/Q$ factor was published in 2012 (Lainey et al. 2012). At the time of their analysis, the authors did not have access to astrometric data from the ISS camera. A set of observations covering the period 1886-2009 was required to obtain a reliable signal from the astrometric data. Lainey et al. (2012) found an intense tidal dissipation ratio ($k_2/Q = (2.3+/-0.7)$ x $10-4$), which was about 10 times higher than the usual value estimated from theoretical arguments (Goldreich and Soter 1966). To support their measurement, the authors noticed that eccentricity equilibrium for Enceladus could now account for the huge heat emitted from Enceladus' south pole (Spencer et al. 2009). Moreover, the measured $k_2/Q$ was found to be poorly sensitive to the tidal frequency, within the uncertainty of the measurement, on the short frequency interval considered. Such a large tidal dissipation within Saturn suggested that the moon current orbits would be inconsistent with the moon formations 4.5 Byr ago above the synchronous orbit in the Saturnian subnebulae. But it would be compatible with a new model of satellite formation in which the Saturnian satellites formed possibly over a longer timescale at the outer edge of the main rings (Charnoz et al. 2011).

A second estimation of Saturnian tides was performed several years later, which benefitted from ISS data and presented an independent analysis of the Saturnian system from both IMCCE-Paris Observatory and JPL (Lainey et al. 2017). The use of ISS-Cassini data significantly reduced the uncertainty on the measurements and revealed a strong dissipative ratio at Rhea's tidal frequency. Such result was interpreted by Fuller et al. (2016) as the sign of a resonance locking mechanism between the moons and internal oscillation modes of the planet which can produce rapid tidal migration (Fuller et al. 2024). Using Rhea's orbital expansion, Fuller et al. (2016) predicted that Titan's orbit could similarly be involved in tidal locking, allowing its orbit to expand extremely fast over time.

### 5.2 Titan's Orbital Expansion

The hypothesis of a fast orbital expansion was tackled by Lainey et al. (2020) using two different approaches. The first one relied on radio-science data acquired during ten of the Titan's flyby by Cassini. Solving for both Cassini and Titan's orbit over a decade, the authors found a peculiar feature on Titan's orbital motion that suggested a strong orbital expansion corresponding to a tidal Saturnian Q of 124 (105-150 at 3-sigma level). In particular, as discussed in Sect. 3.2, the Cassini radio science data accuracy proved fundamental to obtain Titan's orbital acceleration over the period of Cassini mission. The second method relied on classical astrometry data (both from ground and space), adding new images involving Hyperion and Titan. This solution provided a Q estimation of 61 (30-301 at 3-sigma level). Both solutions were found consistent within 2-sigma uncertainties.

In their work, the authors provided a new estimation of tidal dissipation at mid-sized moons frequency, also. The roughly constant migration time $t_{tide}$ for several of Saturn's moons suggested the authors that resonance locking with inertial waves, rather than gravity modes (which predicts smaller $t_{tide}$ values for outer moons), is the most probable explanation for the moons' migration. This also helped to explain why mean-motion resonances between moons have survived, because resonance locking with gravity modes typically results in divergent migration that can disrupt mean-motion resonances between moons, whereas resonance locking with inertial waves does not necessarily produce divergent migration and allows mean-motion resonances to survive.

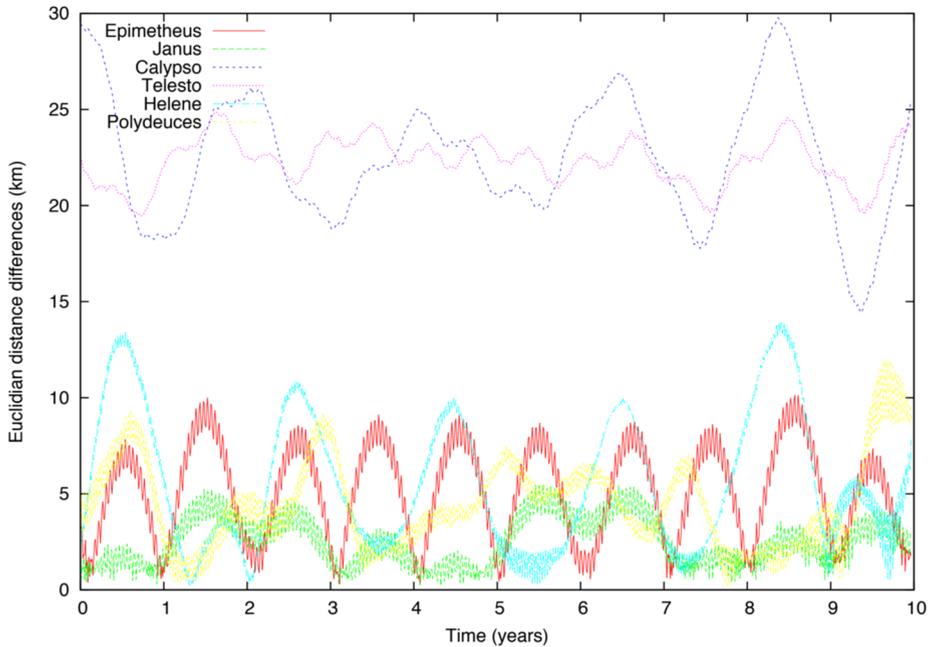

**Fig. 7** Postfit residuals associated with cross-tidal effects from Lainey et al. (2017). Kilometric signals appear associated with the Saturn's tidal bulge at Janus/Epimetheus', Mimas', Enceladus', Tethys' and Dione's tidal frequencies

More recently, the large tidal expansion of Titan was called into question by Jacobson (2022). Using similar dataset, the author could not reproduce the results obtained from radio-science data by Lainey et al. (2020). But the author found that a Voyager 1 flyby in addition to Cassini's flyby data can measure Titan's orbital expansion. Jacobson (2022) obtained a value about ten times larger for the Saturnian tidal Q parameter. Future analysis will have to be done to understand the source of such a discrepancy.

### 5.3 Love Number Estimations

The large number and accuracy of the ISS-Cassini dataset concerned also the small coorbital moons of Tethys and Dione, called Telesto, Calypso, Helen and Polydeuces. In particular, (Lainey et al. 2017) could use the orbital motion of these moons to determine, for the first time from observations, the Love number $k_2$ of Saturn. This determination allowed the authors to discriminate significantly the interior modelling of Saturn. In particular, the Love number appeared to be extremely complementary in term of interior constraints to the usual gravitational coefficients $J_2$, $J_4$ and $J_6$ determined from radio-science data (Lainey et al. 2017).

Figure 7 shows the incompressible dynamical effect of Saturn's $k_2$ on the orbital motion of the system's small moons after fitting the moons' initial conditions, masses, Saturn's $J_2$, polar orientation and precession and Saturn's tidal Q. This represents up to several tens of kilometers on the motion of Telesto and Calypso. Such signal appears extremely clearly in the post-fit residuals. Even more recently, it has been possible to quantify Saturn's $k_2$ at several frequencies. To do this, Lainey et al. (2024) took advantage of a number of observations covering the whole thirteen years of the Cassini mission. They were able to determine the number of Love $k_2$ of Saturn at the frequency of Janus/Epimetheus, Mimas, Tethys and Dione. Within the error bars, Saturn's $k_2$ solution appears consistent with both a static value and predictions from numerical models of the tidal response of Saturn, including contributions from fundamental modes, gravito-inertial modes, and inertial waves (Lainey et al. 2024).

## 6 The Icy Giants: Uranus and Neptune

The first accelerations reported for icy giants is the work of Emelyanov and Nikonchuk (2016). But the large and negative sign makes these estimations extremely suspicious. Considering the large distance of Uranus and Neptune, it is challenging to make any reliable constraints before a dedicated space mission or the release of more Gaia data. More recently, Jacobson and Park (2025) determined orbital expansion of the main Uranian moons implying $Q = 678 \pm 231$ (assuming $k_2 = 0.3$). While their values (including the sign) are significantly different from Emelyanov and Nikonchuk (2016), they still partly use the same dataset. The question of potential biases within the old astrometric data, particularly less accurate than for the Jovian and Saturnian system is raised. Digitisation of old plates may help in the future to solve this question (Sect. 2.1.1).

For the moment, the Gaia data available covers the period 2014-2020. With a magnitude of between 14 and 15 (with the exception of Miranda, which is not observable by Gaia), the main Uranian satellites correspond perfectly to Gaia's capabilities. Along track measurements guarantee fantastic precision on the orbits of these moons, with residuals of typically 1 milliarcsecond (Fig. 8). Recent theoretical estimation of the Uranian tidal parameter Q expect possible values as low as $10^3$ (Nimmo 2023). Lainey (2016) studied the tidal signal amplitude to look for assuming $Q = 100$ for Uranus. The author found a post-fit signal of about 40 mas on Ariel over 100 years of data. Assuming linearity in $Q$, the Nimmo (2023) value implies a post-fit amplitude associated to tides of about few to several mas, only. Limiting the analysis to the currently available Gaia interval data, this would induce a signal even few hundred times smaller, since tides are a quadratic effect over time, way below the Gaia precision. Nevertheless, a $Q$ of several hundred might be detectable with the arrival of the last Gaia astrometric catalogue in late 2020s and the use of the most accurate astrometric data gathered over more than a century, including Voyager (Jacobson 1992) and mutual events data (Arlot et al. 2013).

For the Neptune system, Lainey (2016) provided a similar study, assuming Q=100 for the planet. A signal of 5 mas over 100 years was found. This is almost an order of magnitude smaller than for the Uranian system, due to the largest distance to the Earth and the significant semi-major axis of Triton. A quantification of tides in Neptune may well depend on the existence of a future space mission.

## 7 Exoplanets

With the discovery of 51 Peg b (Mayor and Queloz 1995), a so-called hot-Jupiter on an extremely tight 4-day orbit around its star, tides have been recognized early as crucial in defining the orbital properties of exoplanets: close-in planets, with orbital periods less than about

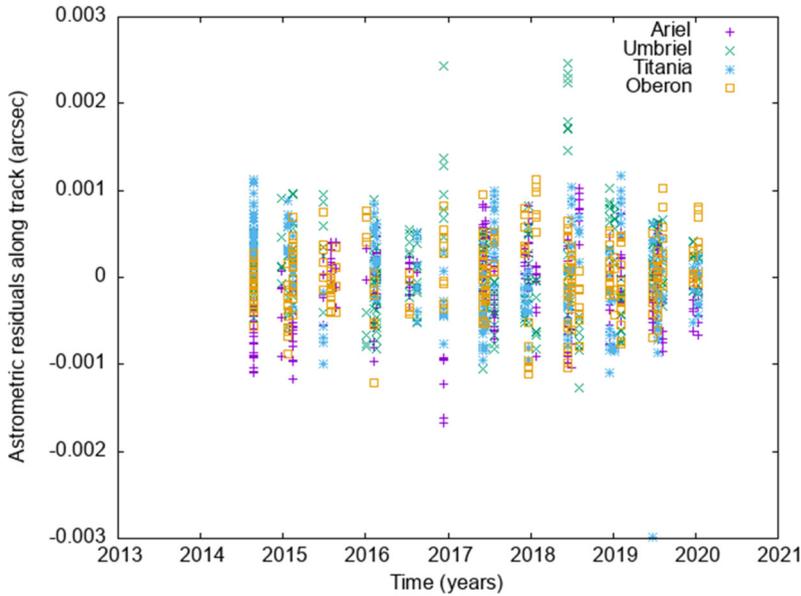

**Fig. 8** Posfit residuals for the main Uranian moons with the Gaia data. Being too close to Uranus, Miranda's data are not available

10 days, should be synchronized with their parent star (Guillot et al. 1996) and circularized (Rasio et al. 1996). Tides have been shown to potentially affect the sizes of hot Jupiters (e.g., Bodenheimer et al. 2001; Guillot and Showman 2002; Guillot and Havel 2011). It was later found that Hot Jupiters generally orbit in the equatorial plane of their star except when orbiting hot stars with no outer convective zone (see Winn and Fabrycky 2015), showing the importance of tides in realigning the orbital plane and of convection in ensuring an efficient dissipation (see Zahn 1977; Ogilvie 2014).

The tidal evolution timescales were derived by Hut (1981), Eggleton et al. (1998) on the basis of an assumed quality factor inversely proportional to the efficiency of dissipation of the tidal energy per orbit. It is useful to consider simple configurations, with a star+planet (« hot Jupiter ») system defined by their masses $M_{1,2}$, radii $R_{1,2}$, modified quality factors $Q'_{1,2}$, semi-major axis a, eccentricity i, orbital period P, stellar spin period $P_\star$. In that case the mean orbital evolution timescale is (Barker and Ogilvie 2009):

$$\tau_a \equiv -\frac{2}{13}\frac{a}{\dot{a}} \approx 12.0 \text{ Myr} \left(\frac{Q'_1}{10^6}\right)\left(\frac{M_1}{M_\odot}\right)^{8/3}\left(\frac{M_2}{M_J}\right)^{-1}\left(\frac{R_1}{R_\odot}\right)^{-5}\left(\frac{P}{1\text{ d}}\right)^{13/3}\left(1-\frac{P}{P_\star}\right)^{-1}.$$

This implies that if, as is the case for most hot Jupiters $P > P_\star$, the planet's orbit will progressively shrink. The $P^{13/3}$ dependence however implies that only the planets with the shortest orbits will be affected in a measurable way.

The orbit is also circularized on a timescale:

$$\tau_e \equiv -\frac{e}{\dot{e}} \approx 16.8 \text{ Myr} \left(\frac{Q'_1}{10^6}\right)\left(\frac{M_1}{M_\odot}\right)^{8/3}\left(\frac{M_2}{M_J}\right)^{-1}\left(\frac{R_1}{R_\odot}\right)^{-5}\left(\frac{P}{1\text{ d}}\right)^{13/3}$$
$$\times \left[\left(f_1(e^2) - \frac{11}{18}\frac{P}{P_\star}f_2(e^2)\right) + \beta\left(f_1(e^2) - \frac{11}{18}f_2(e^2)\right)\right]^{-1},$$

where $f_{1,2}(e^2)$ are functions defined in Barker and Ogilvie (2009) with values close to unity and β defines the ratio of the importance of the planetary to the stellar tides:

$$\beta = \frac{Q'_1}{Q'_2} \left(\frac{M_1}{M_2}\right)^{1/3} \left(\frac{\overline{\rho}_1}{\overline{\rho}_2}\right)^{5/3},$$

and $\overline{\rho}_{1,2}$ are the stellar and planetary mean densities, respectively.[1] Hot Jupiters have mean densities that are generally between 1 and 0.1 g/cm$^3$ for the most inflated ones implying that even though they are ∼1000 times less massive than their parent star, their low density and low quality factor can lead to β > 1, i.e., planetary tides dominating the circularization over the stellar tides.

Lastly, it the orbital and stellar equatorial planes are misaligned, dissipation of the stellar tide (raised on the star by the planet) would align them on a timescale

$$\tau_i \equiv -\frac{i}{di/dt} \approx 70.0 \text{ Myr} \left(\frac{Q'_1}{10^6}\right)\left(\frac{M_1}{M_\odot}\right)\left(\frac{M_2}{M_J}\right)^{-2}\left(\frac{R_1}{R_\odot}\right)^{-3}\left(\frac{P}{1 \text{ d}}\right)^4\left(\frac{\Omega_1}{\Omega_0}\right)$$
$$\times \left[1 - \frac{P}{2P_\star}\left(1 - \frac{1}{\alpha}\right)\right]^{-1},$$

where a circular orbit and small i are assumed, $\Omega_1$ is the star spin frequency, $\Omega_0 = 5.8 \times 10^{-6}$ s$^{-1}$ corresponds to a ∼12.5 d spin period and α is the ratio of orbital to stellar spin angular momentum.

From ultra-precise timing of transits over a long time, at least two planets have been measured to be falling into their star. WASP-12b, an inflated 1.9 R$_J$, 1.4 M$_J$ hot Jupiter with an orbital period of about P=26 hours around a solar-type star, has been measured to have a shrinking orbit with $\dot{P} = -29.81 \pm 0.94$ ms/yr (Maciejewski et al. 2016a,b; Wong et al. 2022), compatible with a stellar quality factor $Q'_\star = 1.5 \times 10^5$, on the lower end of expectations from stellar models. Kepler-1658b, a massive hot Jupiter (1.1 R$_J$, 1.4 M$_J$) with a P=3.85 day orbit around an evolved F-type star (2.9 R$_\odot$, 1.5 M$_\odot$), was found to also have a decreasing orbital period $\dot{P} = -131^{+20}_{-22}$ ms/yr, corresponding to an infall timescale $\tau_a \sim 2.5$ Myr and a stellar quality factor $Q'_\star = 2.5 \times 10^4$ (Vissapragada et al. 2022). This quality factor appears much lower than expectations, except during a very limited ∼100 yrs evolution phase during which inertial waves may be driven in the stellar interior indicating that other explanations should be found (Barker et al. 2024).

Studies of orbital circularization have also been able to put constraints on the tidal quality factors for hot Jupiters as shown in Fig. 8. Studies of individual systems have the potential to provide some constraints: For example, the unique configuration of HAT-P-13b, an eccentric hot Jupiter in resonance with an outer eccentric companion puts constraints on tidal dissipation in the inner planet (Batygin et al. 2009). This is also the case if WASP-12b's rapid inward migration is caused by obliquity tides (Millholland and Laughlin 2018). Lower limits can be obtained from the period decay of XO-3b (Yang and Wei 2022), and from detailed analyses using asteroseimic information for the Kepler-91 system (Fellay et al. 2023).

Studies of ensemble of systems provide powerful statistical constraints. Studies of orbit circularization in open clusters are promising because the known stellar ages lift at least one degeneracy. Figure 8 shows that these studies applied to 0.7 Gyr M67 and 4 Gyr Hyades and Praesepe lead to values of $Q'_p$ ranging from about $2 \times 10^6$ to $2 \times 10^7$ (O'Connor and

---

[1] This equation corrects a typo in Barker and Ogilvie (2009) in which had been inverted $\beta Q'_1/Q'_2$.

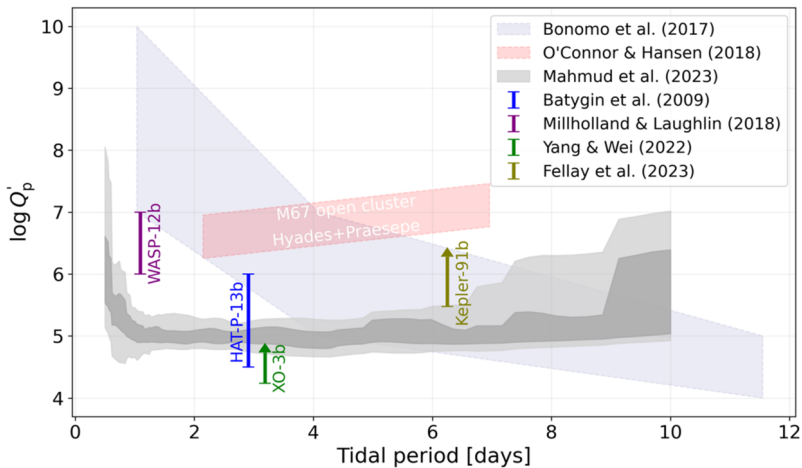

**Fig. 9** Constraints obtained on the modified planetary quality factor $log\, Q'_p$ as a function of the tidal period obtained from different works as labelled. Constraints on individual systems (WASP-12b, HAT-P-13b) and lower limits (XO-3b, Kepler-91b) are indicated with error bars, and arrows, respectively. Upper limits obtained from the circularization of planets supposed to be initially eccentric coupled to lower limits for systems with a non-zero eccentricity are shown in pale blue and approximated from the results of Bonomo et al. (2017). Constraints obtained from 6 planets in the open clusters M67, Hyades and Praesepe by O'Connor & Hansen and recalculated as a function of $log\, Q'_p$ are shown as a pink area. Constraints derived from tidal evolution calculations of 78 planets analyzed with a Bayesian approach (Mahmud et al. 2023) are shown in gray

Hansen 2018), but the small number of planets (6) in these clusters limits the robustness of the estimation. With a selected sample of 211 planets, Bonomo et al. (2017) focused on deriving upper limits on $Q'_p$ for planets with circular orbits and lower limits for planets with known eccentricites yielding constraints that appear to favor higher values of $Q'_p$ for the shortest orbits and lower ones, down to about 104 for periods of 10 days and higher. Perhaps the most involved study so far is that of Mahmud et al. (2023) which includes a detailed study of the combined tidal evolution of the star and planet for a sample of 78 single-planet systems with the hypothesis that the planet quality factor only depends on the tidal period. The Bayesian analysis results in rather well constrained values $Q'_p = 10^{5\pm 0.5}$, independently of the tidal period between 1 and 10 days. The discrepancies between the different methods seen in Fig. 8 however shows that more work is needed.

## 8 Conclusion

The determination of tidal parameters within giant planet systems has taken a major leap forward in the last fifteen years, thanks in particular to data from the Cassini and Juno space probes. The imminent arrival of the new generation of probes such as Juice (ESA) and Europa Clipper (NASA) should take us a step further in our understanding of the dissipative phenomena at work within these systems. The question of the origin of the high energy dissipation under the effect of tidal phenomena, as well as its dependence on the excitation frequency, remains a subject for further study if we ultimately wish to understand the past dynamic evolution of these systems.

Beyond the Solar system, data from future ground- and space-based programs such as PLATO (ESA) should enable us to balance what we observe today in the Solar system with the dynamic richness of the thousands of exoplanetary systems known today. It is to be hoped that the decade to come will enable us to integrate the dynamic study of all these systems into a global context, clearly revealing the physical mechanisms underlying tidal dissipation in these objects.

**Acknowledgements** This paper resulted from the Workshop "New Vision of the Saturnian System in the Context of a Highly Dissipative Saturn" held at the International Space Science Institute (ISSI) in Bern (Switzerland).

**Funding Information** Open access funding provided by Alma Mater Studiorum - Università di Bologna within the CRUI-CARE Agreement. M. Zannoni is grateful to the Italian Space Agency (ASI) for financial support through Agreement No. 2023-6-HH.O.

## Declarations

**Competing Interests** The authors have no conflicts of interest to report.

## References

Aksnes K, Franklin FA (2001) Secular acceleration of Io derived from mutual satellite events. Astron J 122:2734–2739

Arlot J-E, et al (2013) Astrometric results of observations of mutual occultations and eclipses of the Uranian satellites in 2007. Astron Astrophys 557

Asmar SW, Armstrong JW, Iess L, Tortora P (2005) Spacecraft Doppler tracking: noise budget and accuracy achievable in precision radio science observations. Radio Sci 40(2):1–9. https://doi.org/10.1029/2004RS003101

Avdyushev VA (2004) A new system of initial parameters for numerical simulation of the motion of Jupiter's Galilean satellites. Sol Syst Res 38:238–240

Barker AJ, Ogilvie GI (2009) On the tidal evolution of hot Jupiters on inclined orbits. Mon Not R Astron Soc 395:2268

Barker AJ, Efroimsky M, Makarov VV, Veras D (2024) On the orbital decay of the gas giant Kepler-1658b. Mon Not R Astron Soc 527:5131

Batygin K, Bodenheimer P, Laughlin G (2009) Determination of the interior structure of transiting planets in multiple-planet systems. Astrophys J 704:L49

Bertotti B, Comoretto G, Iess L (1993) Doppler tracking of spacecraft with multi-frequency links. Astron Astrophys 269:608–616

Bodenheimer P, Lin DNC, Mardling RA (2001) On the tidal inflation of short-period extrasolar planets. Astrophys J 548:466

Bonomo AS, Desidera S, Benatti S, Borsa F, Crespi S, et al (2017) The GAPS programme with HARPS-N at TNG. XIV. Investigating giant planet migration history via improved eccentricity and mass determination for 231 transiting planets. Astron Astrophys 602:A107

Brozovic M, et ak (2020) Arecibo Radar Astrometry of the Galilean Satellites from 1999 to 2016. Astron J 159

Buccino DR, Kahan DS, Parisi M, Paik M, Barbinis E, Yang O, Park RS, Tanner A, Bryant SH, Jongeling A (2021) Performance of Earth troposphere calibration measurements with the advanced water vapor radiometer for the Juno gravity science investigation. Radio Sci 56(12):1–9. https://doi.org/10.1029/2021RS007387

Campbell JK, Anderson JD (1989) Gravity field of the Saturnian system from Pioneer and Voyager tracking data. Astron J 97:1485. https://doi.org/10.1086/115088

Charnoz S, et al (2011) Accretion of Saturn's mid-sized moons during the viscous spreading of young massive rings: solving the paradox of silicate-poor rings versus silicate-rich moons. Icarus 216(2):535–550. https://doi.org/10.1016/j.icarus.2011.09.017

Cooper NJ, Murray CD, Lainey V, Tajeddine R, Evans MW, Williams GA (2014) Cassini ISS mutual event astrometry of the mid-sized Saturnian satellites 2005-2012. Astron Astrophys 572

Cooper NJ, et al (2018) The Caviar software package for the astrometric reduction of Cassini ISS images: description and examples. Astron Astrophys 610

de Sitter W (1916) Jupiter, the longitude of the satellites of, from photographs taken at the Cape Observatory in 1913. Mon Not R Astron Soc 76:448

de Sitter W (1928) Orbital elements determining the longitudes of Jupiter's satellites, derived from observations. Leiden Ann 16(2):1–92

Desage L, Hérique A, Lainey V, Kofman W, Cicchetti A, Orosei R (2024) MARSIS data as a new constraint for the orbit of Phobos. Astron Astrophys 686:A136. https://doi.org/10.1051/0004-6361/202348655

Dewberry JW (2023) Dynamical tides in Jupiter and other rotationally flattened planets and stars with stable stratification. Mon Not R Astron Soc 521(4):5991–6004. https://doi.org/10.1093/mnras/stad546

Dewberry JW, Lai D (2022) Dynamical tidal Love numbers of rapidly rotating planets and stars. Astrophys J 925(2):124. https://doi.org/10.3847/1538-4357/ac3ede

Dourneau G (1987) Observations et études du mouvement des huit principaux satellites de Saturne. Thèse, Bordeaux

Durante D, Parisi M, Serra D, Zannoni M, Notaro V, Racioppa P, Buccino DR, Lari G, Gomez Casajus L, Iess L, Folkner WM, Tommei G, Tortora P, Bolton SJ (2020) Jupiter's gravity field halfway through the Juno mission. Geophys Res Lett 47(4):1–8. https://doi.org/10.1029/2019GL086572

Durante D, Guillot T, Iess L, Stevenson DJ, Mankovich CR, Markham S, Galanti E, Kaspi Y, Zannoni M, Gomez Casajus L, Lari G, Parisi M, Buccino DR, Park RS, Bolton SJ (2022) Juno spacecraft gravity measurements provide evidence for normal modes of Jupiter. Nat Commun 13(1):4632. https://doi.org/10.1038/s41467-022-32299-9

Duriez L, Vienne A (1991) A general theory of motion for the eight major satellites of Saturn. I. Equations and method of resolution. Astron Astrophys 243:263–275

Eggleton PP, Kiseleva LG, Hut P (1998) The equilibrium tide model for tidal friction. Astrophys J 499:853

Emelyanov NV (2017) Current problems of dynamics of moons of planets and binary asteroids based on observations. Sol Syst Res 51:20–37

Emelyanov NV, Nikonchuk DV (2013) Ephemerides of the main Uranian satellites. Mon Not R Astron Soc 436:3668–3679. https://doi.org/10.1093/mnras/stt1851

Emelyanov NV, Nikonchuk DV (2016) Erratum: Ephemerides of the main Uranian satellites: Table 1. Mon Not R Astron Soc 461:1718

Emelyanov NV, Samorodov MY (2015) Analytical theory of motion and new ephemeris of Triton from observations. Mon Not R Astron Soc 454:2205–2215. https://doi.org/10.1093/mnras/stv2116

Fayolle M, Magnanini A, Lainey V, Dirkx D, Zannoni M, Tortora P (2023) Combining astrometry and JUICE – Europa Clipper radio science to improve the ephemerides of the Galilean moons. Astron Astrophys 677:A42. https://doi.org/10.1051/0004-6361/202347065

Fellay L, Pezzotti C, Buldgen G, Eggenberger P, Bolmont E (2023) Constraints on planetary tidal dissipation from a detailed study of Kepler 91b. Astron Astrophys 669:A2

Folkner WM, Williams JG, Boggs DH, Park RS, Kuchynka P (2014) The Planetary and Lunar Ephemerides DE430 and DE431. The Interplanetary Network Progress Report, 1–81. http://tmo.jpl.nasa.gov/progress_report/42-196/196C.pdf

Fuller J, Luan J, Quataert E (2016) Resonance locking as the source of rapid tidal migration in the Jupiter and Saturn moon systems. Mon Not R Astron Soc 458:3867–3879

Fuller J, Guillot T, Mathis S, Murray C (2024) Tidal Dissipation in Giant Planets. Space Sci Rev 220

Gaia Collaboration (2023) Gaia data release 3: summary of the content and survey properties. Astron Astrophys 674:A1. https://doi.org/10.1051/0004-6361/202243940

Gavrilov SV, Zharkov VN (1977) Love numbers of the giant planets. Icarus 32:443–449. https://doi.org/10.1016/0019-1035(77)90015-X

Goldreich P, Soter S (1966) Q in the Solar System. Icarus 5(1–6):375–389. https://doi.org/10.1016/0019-1035(66)90051-0

Goldstein SJ Jr, Jacobs KC (1995) A recalculation of the secular acceleration of Io. Astron J 110:3054–3057

Gomez Casajus L (2019) Development of methods for the global ephemerides estimation of the gas giant satellite systems. PhD thesis, Alma Mater Studiorum Università di Bologna

Gomez Casajus L, Ermakov AI, Zannoni M, Keane JT, Stevenson DJ, Buccino DR, Durante D, Parisi M, Park RS, Tortora P, Bolton SJ (2022) Gravity field of Ganymede after the Juno extended mission. Geophys Res Lett 49(24):1–10. https://doi.org/10.1029/2022GL099475

Guillot T, Havel M (2011) An analysis of the CoRoT-2 system: a young spotted star and its inflated giant planet. Astron Astrophys 527:A20

Guillot T, Showman AP (2002) Evolution of "51 Pegasus b-like" planets. Astron Astrophys 385:156

Guillot T, Burrows A, Hubbard WB, Lunine JI, Saumon D (1996) Giant planets at small orbital distances. Astrophys J 459:L35

Guillot T, Miguel Y, Militzer B, Hubbard WB, Kaspi Y, Galanti E, Cao H, Helled R, Wahl SM, Iess L, Folkner WM, Stevenson DJ, Lunine JI, Reese DR, Biekman A, Parisi M, Durante D, Connerney JEP, Levin SM, Bolton SJ (2018) A suppression of differential rotation in Jupiter's deep interior. Nature 555(7695):227–230. https://doi.org/10.1038/nature25775

Harmon JK, Ostro SJ, Chandler JF, Hudson RS (1994) Radar ranging to Ganymede and Callisto. Astron J 107:1175

Harper D, Taylor DB (1993) The orbits of the major satellites of Saturn. Astron Astrophys 268:326–349

Harper D, Taylor DB, Sinclair AT, Kaixian S (1988) The theory of the motion of Iapetus. Astron Astrophys 191:381–384

Helled R, Stevenson DJ, Lunine JI, Bolton SJ, Nettelmann N, Atreya S, Guillot T, Militzer B, Miguel Y, Hubbard WB (2022) Revelations on Jupiter's formation, evolution and interior: challenges from Juno results. In: Icarus, vol 378. Academic Press, San Diego. https://doi.org/10.1016/j.icarus.2022.114937

Howell SB (2006) Handbook of CCD astronomy. Cambridge University Press, Cambridge

Hut P (1981) Tidal evolution in close binary systems. Astron Astrophys 99:126

Idini B, Stevenson DJ (2021) Dynamical tides in Jupiter as revealed by Juno. Planet Sci J 2(2):69. https://doi.org/10.3847/psj/abe715

Idini B, Stevenson DJ (2022a) The Lost Meaning of Jupiter's High-degree Love Numbers. Planet Sci J 3(1). https://doi.org/10.3847/PSJ/ac4248

Idini B, Stevenson DJ (2022b) The Gravitational Imprint of an Interior–Orbital Resonance in Jupiter–Io. Planet Sci J 3(4). https://doi.org/10.3847/PSJ/ac6179

Iess L, di Benedetto M, James N, Mercolino M, Simone L, Tortora P (2014) Astra: interdisciplinary study on enhancement of the end-to-end accuracy for spacecraft tracking techniques. Acta Astronaut 94(2):699–707. https://doi.org/10.1016/j.actaastro.2013.06.011

Iess L, Folkner WM, Durante D, Parisi M, Kaspi Y, Galanti E, Guillot T, Hubbard WB, Stevenson DJ, Anderson JD, Buccino DR, Gomez Casajus L, Milani A, Park RS, Racioppa P, Serra D, Tortora P, Zannoni M, Cao H, Bolton SJ, et al (2018) Measurement of Jupiter's asymmetric gravity field. Nature 555(7695). https://doi.org/10.1038/nature25776

Iess L, Militzer B, Kaspi Y, Nicholson P, Durante D, Racioppa P, Anabtawi A, Galanti E, Hubbard WB, Mariani MJ, Tortora P, Wahl SM, Zannoni M (2019) Measurement and implications of Saturn's gravity field and ring mass. Science 364(6445):eaat2965. https://doi.org/10.1126/science.aat2965

Iess L, Asmar SW, Cappuccio P, Cascioli G, de Marchi F, di Stefano I, Genova A, Ashby N, Barriot J, Bender P, Benedetto C, Border JS, Budnik F, Ciarcia S, Damour T, Dehant V, di Achille G, di Ruscio A, Fienga A, Zannoni M, et al (2021) Gravity, geodesy and fundamental physics with BepiColombo's MORE investigation. Space Sci Rev 217(1):21. https://doi.org/10.1007/s11214-021-00800-3

Iess L, et al (2024) The 3GM investigation of ESA's JUICE mission: Geodesy and Geophysics of Jupiter and the Galilean Moons. Space Sci Rev. Under submission

Jacobson RA (1992) Astrographic observations of the major Uranian satellites from Voyager 2. Astron Astrophys Suppl Ser 96:549–563

Jacobson RA (2001) The gravity field of the Jovian system and the orbits of the regular Jovian satellites. In: AAS/Division for Planetary Sciences Meeting, vol 33

Jacobson RA (2004) The orbits of the major saturnian satellites and the gravity field of Saturn from spacecraft and Earth-based observations. Astron J 128(1):492–501. https://doi.org/10.1086/421738

Jacobson RA (2014) The orbits of the Uranian satellites and rings, the gravity field of the Uranian system, and the orientation of the pole of Uranus. Astron J 148:76. https://doi.org/10.1088/0004-6256/148/5/76

Jacobson RA (2022) The orbits of the main Saturnian satellites, the Saturnian system gravity field, and the orientation of Saturn's pole. Astron J 164:199. https://doi.org/10.3847/1538-3881/ac90c9

Jacobson RA, Park RS (2025) The orbits of Uranus, its satellites and rings, the gravity field of the Uranian system, and the orientation of the poles of Uranus and its satellites. Astron J 169:65. https://doi.org/10.3847/1538-3881/ad99d1

Jacobson RA, Riedel JE, Taylor AH (1991) The orbits of Triton and Nereid from spacecraft and earthbased observations. Astron Astrophys 247:565–575

Jacobson RA, Haw RJ, McElrath TP, Antreasian PG (2000) A comprehensive orbit reconstruction for the Galileo prime mission in the J2000 system. J Astronaut Sci 48(4):495–516. https://doi.org/10.1007/bf03546268

Jacobson RA, Antreasian PG, Bordi JJ, Criddle KE, Ionasescu R, Jones JB, Mackenzie RA, Meek MC, Parcher DW, Pelletier FJ, Owen WM, Roth DC, Roundhill IM, Stauch JR (2006) The gravity field of the saturnian system from satellite observations and spacecraft tracking data. Astron J 132(6):2520–2526. https://doi.org/10.1086/508812

Kaspi Y, Galanti E, Hubbard WB, Stevenson DJ, Bolton SJ, Iess L, Guillot T, Bloxham J, Connerney JEP, Cao H, Durante D, Folkner WM, Helled R, Ingersoll AP, Levin SM, Lunine JI, Miguel Y, Militzer B, Parisi M, Wahl SM (2018) Jupiter's atmospheric jet streams extend thousands of kilometres deep. Nature 555(7695):223–226. https://doi.org/10.1038/nature25793

Kozai Y (1957) On the astronomical constants of Saturnian satellites system. Ann Tokyo Astron Obs 5:73–106

Lainey V (2008) A new dynamical model for the Uranian satellites. Planet Space Sci 56:1766–1772. https://doi.org/10.1016/j.pss.2008.02.015

Lainey V (2016) Quantification of tidal parameters from Solar System data. Celest Mech Dyn Astron 126:145–156. https://doi.org/10.1007/s10569-016-9695-y

Lainey V, Duriez L, Vienne A (2004) New accurate ephemerides for the Galilean satellites of Jupiter. I. Numerical integration of elaborated equations of motion. Astron Astrophys 420:1171–1183

Lainey V, Arlot J-E, Karatekin Ö, van Hoolst T (2009) Strong tidal dissipation in Io and Jupiter from astrometric observations. Nature 459:957–959

Lainey V, et al (2012) Strong tidal dissipation in Saturn and constraints on Enceladus' thermal state from astrometry. Astrophys J 752:14. https://doi.org/10.1088/0004-637X/752/1/14

Lainey V, et al (2017) New constraints on Saturn's interior from Cassini astrometric data. Icarus 281:286–296

Lainey V, et al (2020) Resonance locking in giant planets indicated by the rapid orbital expansion of Titan. Nat Astron 4:1053–1058

Lainey V, Dewberry W, Fuller J, Cooper N, Rambaux N, Zhang Q (2024) Tidal frequency dependence of the Saturnian $k_2$ Love number. Astron Astrophys 684:L3. https://doi.org/10.1051/0004-6361/202449639

Lasagni Manghi R, Zannoni M, Tortora P, Martellucci A, de Vicente J, Villalvilla J, Mercolino M, Maschwitz G, Rose T (2021) Performance Characterization of ESA's Tropospheric Delay Calibration System for Advanced Radio Science Experiments. Radio Sci 56(10). https://doi.org/10.1029/2021RS007330

Lasagni Manghi R, Bernacchia D, Gomez Casajus L, Zannoni M, Tortora P, Martellucci A, de Vicente J, Villalvilla J, Maschwitz G, Cappuccio P, Iess L (2023) Tropospheric delay calibration system performance during the first two BepiColombo solar conjunctions. Radio Sci 58(2):1–15. https://doi.org/10.1029/2022RS007614

Laskar J (1986) A general theory for the Uranian satellites. Astron Astrophys 166:349–358

Lieske JH (1974) A method of revitalizing Sampson's theory of the Galilean satellites. Astron Astrophys 31:137

Lieske JH (1986b) A collection of Galilean satellite eclipse observations, 1652-1983. II. Astron Astrophys Suppl Ser 63:143–202

Lieske JH (1986a) A collection of Galilean satellite eclipse observations, 1652-1983. I. Astron Astrophys 154:61–76

Lieske JH (1987) Galilean satellite evolution - observational evidence for secular changes in mean motions. Astron Astrophys 176:146–158

Lieske JH (1996) Galilean satellites and the Galileo space mission. Celest Mech Dyn Astron 66:13–20

Lin FR, Peng JH, Zheng ZJ, Peng QY (2019) Characterization of the precision premium in astrometry. Mon Not R Astron Soc 490(3):4382–4387. https://doi.org/10.1093/mnras/stz2871

Lin Y (2023) Dynamical tides in Jupiter and the role of interior structure. Astron Astrophys 671:A37. https://doi.org/10.1051/0004-6361/202245112

Maciejewski G, Dimitrov D, Fernández M, Sota A, Nowak G, et al (2016a) Departure from the constant-period ephemeris for the transiting exoplanet WASP-12. Astron Astrophys 588:L6

Maciejewski G, Dimitrov D, Mancini L, Southworth J, Ciceri S, et al (2016b) New transit observations for HAT-P-30 b, HAT-P-37 b, TrES-5 b, WASP-28 b, WASP-36 b and WASP-39 b. Acta Astron 66:55

Magnanini A, Zannoni M, Casajus LG, Tortora P, Lainey V, Mazarico E, Park RS, Iess L (2024) Joint analysis of JUICE and Europa Clipper tracking data to study the Jovian system ephemerides and dissipative parameters. Astron Astrophys 687:A132. https://doi.org/10.1051/0004-6361/202347616

Mahmud MM, Penev KM, Schussler JA (2023) Measuring tidal dissipation in giant planets from tidal circularization. Mon Not R Astron Soc 525:876

Mallama A, Stockdale C, Krobusek BA, Nelson P (2010) Assessment of the resonant perturbation errors in Galilean satellite ephemerides using precisely measured eclipse times. Icarus 210:346–357

Mariotti G, Tortora P (2013) Experimental validation of a dual-uplink multifrequency dispersive noise calibration scheme for deep space tracking. Radio Sci 48. https://doi.org/10.1002/rds.20024

Markham S, Durante D, Iess L, Stevenson D (2020) Possible evidence of p-modes in Cassini measurements of Saturn's gravity field. Planet Sci J 1:27. https://doi.org/10.3847/PSJ/ab9f21

Mayor M, Queloz D (1995) A Jupiter-mass companion to a solar-type star. Nature 378:355

Mazarico E, Buccino DR, Castillo-Rogez JC, Dombard AJ, Genova A, Hussmann H, Kiefer WS, Lunine JI, McKinnon WB, Nimmo F, Park RS, Roberts JH, Srinivasan DK, Steinbrügge G, Tortora P, Withers P (2023) The Europa Clipper gravity and radio science investigation. Space Sci Rev 219(4):30. https://doi.org/10.1007/s11214-023-00972-0

Miguel Y, Bazot M, Guillot T, Howard S, Galanti E, Kaspi Y, Hubbard WB, Militzer B, Helled R, Atreya SK, Connerney JEP, Durante D, Kulowski L, Lunine JI, Stevenson D, Bolton S (2022) Jupiter's inhomogeneous envelope. Astron Astrophys 662:A18. https://doi.org/10.1051/0004-6361/202243207

Milani A, Gronchi G (2009) Theory of orbit determination, vol 1. Cambridge University Press, Cambridge. https://doi.org/10.1017/CBO9781139175371

Militzer B, Hubbard WB, Wahl S, Lunine JI, Galanti E, Kaspi Y, Miguel Y, Guillot T, Moore KM, Parisi M, Connerney JEP, Helled R, Cao H, Mankovich C, Stevenson DJ, Park RS, Wong M, Atreya SK, Anderson J, Bolton SJ (2022) Juno spacecraft measurements of Jupiter's gravity imply a dilute core. Planet Sci J 3(8):185. https://doi.org/10.3847/PSJ/ac7ec8

Millholland S, Laughlin G (2018) Obliquity tides may drive WASP-12b's rapid orbital decay. Astrophys J 869:L15

Morgado B, Assafin M, Vieira-Martins R, Camargo JIB, Dias-Oliveira A, Gomes-Jùnior AR (2016) Astrometry of mutual approximations between natural satellites. Application to the Galilean moons. Mon Not R Astron Soc 460:4086–4097

Morgado B, et al (2019) First stellar occultation by the Galilean moon Europa and upcoming events between 2019 and 2021. Astron Astrophys 626:L4. https://doi.org/10.1051/0004-6361/201935500

Nimmo F (2023) Strong tidal dissipation at Uranus? Planet Sci J 4:241. https://doi.org/10.3847/PSJ/ad0cfb

Notaro V, Durante D, Iess L (2019) On the determination of Jupiter's satellite-dependent Love numbers from Juno gravity data. Planet Space Sci 175:34–40. https://doi.org/10.1016/j.pss.2019.06.001

Null GW, Lau EL, Biller ED, Anderson JD (1981) Saturn gravity results obtained from Pioneer 11 tracking data and Earth-based Saturn satellite data. Astron J 86:456. https://doi.org/10.1086/112905

O'Connor CE, Hansen BMS (2018) Constraining planetary migration and tidal dissipation with coeval hot Jupiters. Mon Not R Astron Soc 477:175

Ogilvie GI (2014) Tidal dissipation in stars and giant planets. Annu Rev Astron Astrophys 52:171

Ogilvie GI, Lin DNC (2004) Tidal dissipation in rotating giant planets. Astrophys J 610:477–509. https://doi.org/10.1086/421454

Ojakangas GW, Stevenson DJ (1986) Episodic volcanism of tidally heated satellites with application to Io. Icarus 66:341–358

Owen WM Jr (2003) Cassini ISS Geometric Calibration of April 2003. JPL IOM 312.E-2003

Parisi M, Kaspi Y, Galanti E, Durante D, Bolton SJ, Levin SM, Buccino DR, Fletcher LN, Folkner WM, Guillot T, Helled R, Iess L, Li C, Oudrhiri K, Wong MH (2021) The depth of Jupiter's Great Red Spot constrained by Juno gravity overflights. Science 374(6570):964–968. https://doi.org/10.1126/science.abf1396

Park RS, Folkner WM, Williams JG, Boggs DH (2021) The JPL planetary and lunar ephemerides DE440 and DE441. Astron J 161(3):105. https://doi.org/10.3847/1538-3881/abd414

Park RS, Jacobson RA, Gomez Casajus L, Nimmo F, Ermakov AI, Keane JT, et al (2025) Io's tidal response precludes a shallow magma ocean. Nature 638(8049):69–73. https://doi.org/10.1038/s41586-024-08442-5

Pascu D (1994) An appraisal of the USNO program for photographic astrometry of bright planetary satellites. Galactic and Solar System Optical Astrometry, Proceedings of the RGO / Institute of Astronomy Workshop, Cambridge, 21. Cambridge Univ Press, Cambridge

Peters CF (1981) Numerical integration of the satellites of the outer planets. Astron Astrophys 104:37–41

Porco CC, West RA, Squyres S, et al (2004) Cassini imaging science: instrument characteristics and anticipated scientific investigations at Saturn. Space Sci Rev 115:363–497. https://doi.org/10.1007/s11214-004-1456-7

Rasio FA, Tout CA, Lubow SH, Livio M (1996) Tidal decay of close planetary orbits. Astrophys J 470:1187

Robert V, Desmars J, Lainey V, Arlot JE, Perlbarg AC, Horville D, Aboudarham J, Etienne C, Gérard J, Ilovaisky S, Khovritchev MY, Le Poncin-Lafitte C, Le Van Suu A, Neiner C, Pascu D, Poirier L, Schneider J, Tanga P, Valls-Gabaud D (2021) The NAROO digitization center - overview and scientific program. Astron Astrophys 652:A3. https://doi.org/10.1051/0004-6361/202140472

Sampson RA (1921) Theory of the four great satellites of Jupiter. Mem R Astron Soc 63:1

Serra D, Lari G, Tommei G, Durante D, Casajus LG, Notaro V, Zannoni M, Iess L, Tortora P, Bolton SJ (2019) A solution of Jupiter's gravitational field from Juno data with the ORBIT14 software. Mon Not R Astron Soc 490(1):766–772. https://doi.org/10.1093/mnras/stz2657

Sinclair AT (1977) The orbits of Tethys, Dione, Rhea, Titan and Iapetus. Mon Not R Astron Soc 180:447–459. https://doi.org/10.1093/mnras/180.3.447

Stevenson DJ (2020) Jupiter's interior as revealed by Juno. Annu Rev Earth Planet Sci 48(1):465–489. https://doi.org/10.1146/annurev-earth-081619-052855

Tajeddine R, Cooper NJ, Lainey V, Charnoz S, Murray CD (2013) Astrometric reduction of Cassini ISS images of the Saturnian satellites Mimas and Enceladus. Astron Astrophys 551:A129

Tanga P, et al (2023) Gaia data release 3. The Solar System survey. Astron Astrophys 674:A12

Van Hoolst T, Tobie G, Vallat C, et al (2024) Geophysical characterization of the interiors of Ganymede, Callisto and Europa by ESA's JUpiter ICy moons Explorer. Space Sci Rev 220:54. https://doi.org/10.1007/s11214-024-01085-y

Vasundhara R, Arlot J-E, Descamps P (1996) Dynamics, ephemerides and astrometry of the Solar System. In: Ferraz-Mello S, Morando B, Arlot J-E (eds) Proc. 172nd symp. Int. Astron. Union, IAU, pp 145–149

Vasundhara R, Arlot J-E, Lainey V, Thuillot W (2003) Astrometry from mutual events of Jovian satellites in 1997. Astron Astrophys 410:337–341

Vienne A, Duriez L (1991) A general theory of motion for the eight major satellites of Saturn. II. Short-period perturbations. Astron Astrophys 246:619–633

Vienne A, Duriez L (1992) A general theory of motion for the eight major satellites of Saturn. III. Long-period perturbations. Astron Astrophys 257:331–352

Vienne A, Sarlat JM, Duriez L (1992) About the secular acceleration of mimas. In: Chaos, resonance, and collective dynamical phenomena in the Solar System, IAU Symposium, vol 152. Kluwer Academic, Dordrecht, p 219

Vissapragada S, Chontos A, Greklek-McKeon M, Knutson HA, Dai F, et al (2022) The possible tidal demise of Kepler's first planetary system. Astrophys J 941:L31

Wahl SM, Parisi M, Folkner WM, Hubbard WB, Militzer B (2020) Equilibrium tidal response of Jupiter: detectability by the Juno spacecraft. Astrophys J 891(1):42. https://doi.org/10.3847/1538-4357/ab6cf9

Winn JN, Fabrycky DC (2015) The occurrence and architecture of exoplanetary systems. Annu Rev Astron Astrophys 53:409

Wong I, Shporer A, Vissapragada S, Greklek-McKeon M, Knutson HA, et al (2022) TESS revisits WASP-12: updated orbital decay rate and constraints on atmospheric variability. Astron J 163:175

Yang F, Wei X (2022) Transit timing variation of XO-3b: evidence for tidal evolution of hot Jupiter with high eccentricity. Publ Astron Soc Pac 134:024401

Zahn J (1977) Tidal friction in close binary systems. Astron Astrophys 57:383

Zhang QF, et al (2021) A comparison of centring algorithms in the astrometry of Cassini imaging science subsystem images and Anthe's astrometric reduction. Mon Not R Astron Soc 505:5253–5259

**Publisher's Note** Springer Nature remains neutral with regard to jurisdictional claims in published maps and institutional affiliations.